# Information Cocoons on Social Media:

# Why and How Should the Government Regulate Algorithms

Wen Yang

The Hong Kong University of Science and Technology

wyangbl@connect.ust.hk

# Content

**Abstract**

Information cocoons are frequently cited in the literature on whether and how social media might lead to ideological segregation and political polarization. From the behavioural and communication perspectives, this paper first examines why algorithm-based social media, as opposed to its traditional counterpart, is more likely to produce information cocoons. We then explore populism and short-termism in voting, bias and noise in decision-making, and prerequisite capital for innovation, demonstrating the importance of information diversity for a sustainable information environment. Finally, this study argues for libertarian paternalism by evaluating the criteria and trade-offs involved in regulating algorithms and proposes to employ nudges to address the core issues while preserving freedom of choice.

**Introduction**

Marshall McLuhan stated that "we shape our tools, and thereafter our tools shape us". As a media theorist, the specific tools that McLuhan focused on is the ever-evolving information communication technologies (ICTs) that have the potential to not only reshape the forms and meanings of content, but also to rewire our perception and cognition (McLuhan & Fiore, 1967). One evidence supporting this argument is the printing press invented by Guttenberg which caused an explosion of information production and dissemination since the 15$^{th}$ century. And the resulting boost in literacy rates partly contributed to the subsequent two industrial revolutions (Haldane, 2015). Nevertheless, with almost 99% of the whole amount of information ever created generated in the 21$^{st}$ century, the transforming power of the information revolution considerably outstrips McLuhan's original predictions.

The Internet, search engines and social media indeed have all provided substantial benefits by offering near-limitless access to knowledge and communication at close-to-zero marginal cost. However, as information theorist Herbert Simon argued, an information-rich society may be attention-poor and patience-thin (Simon, 1971). In this regard, the algorithm-based social media have also imposed considerable costs to the society. One of the most frequently cited phenomena in literature on whether social media would facilitate or mitigate ideological segregation and political polarization is information cocoons (Sunstein, 2006), with other similar concepts including the echo chamber effect and the filter bubbles (Cinelli et al., 2021; Pariser, 2011b).

Among the occurrences, the Brexit referendum and the presidency of Donald Trump in 2016 are often used as case studies of social media and information cocoons (Bastos et al., 2018; DiFranzo & Gloria-Garcia, 2017; Geschke et al., 2019; Napoli, 2018; Pariser, 2011a; Su, 2022). And the main concerns revolved around whether information cocoons exist on social media and whether the cocoons constitute a valid variable in the voting outcomes. Even though opposing viewpoints and competing evidence are constantly emerging in this literature, this paper will simply admit the existence of information cocoons on social media, eschewing quantitative measuring and analysis, but rather employ qualitative methods to investigate why algorithm-based social media, as opposed to its traditional counterpart, are more likely to produce cocoons from the behavioural, cognitive and communication perspectives.

Based on these contexts, we then discuss populism and short-termism in voting, bias and noise in decision-making, and prerequisite capitals for innovation as policy problems to make the case for government interventions in social media and algorithm regulations, as well as advocating for information diversity in order to promote long-term social growth and sustainable development. Finally, in terms of policy recommendations, by considering and analyzing the criteria and trade-offs involved in regulating social media algorithms, this paper argues for a form of libertarian paternalism and proposes to use nudges and choice-architecture while respecting the freedom of choice among citizens.

# 1. Policy Context: Social media as brain-reshaping machine

**Blog triumphalism vs Media professionalism**

Since its inception, social media has outperformed traditional media outlets in terms of convenience, immediacy, and cost-effectiveness, eventually becoming the primary news source. According to a study (Shearer et al., 2015), the majorities of Twitter (63%) and Facebook users (63%) said each platform served as a source for news. With more than 1.55 million active users, 968 million people logged onto Facebook daily in June 2015, sharing over 25 billion web articles. This tendency was once referred to as blog triumphalism (Hewitt, 2006), that is, professional journalists of traditional media platforms would become less competitive against grassroot blog reporters. And since 2016, social media has been increasingly recognized as one of the significant variables that influenced the global political dynamics (Lewandowsky et al., 2017; Su, 2022).

Meanwhile, with pervasive application of machine learning and recommendation systems, social media has also played a consequential role in framing the agendas and disseminating information. Decades ago, the MIT Media Lab founder (Negroponte et al., 1997) envisioned a customized medium ("Daily Me") catering to individuals' personal tastes. In this sense, social media algorithms have fulfilled the "Daily Me" prophecy and are catering to different audiences in the digital public sphere by gathering highly relevant topics and like-minded opinions while filtering out divergent content, usually without any substantial engagement of media editors and professionals serving as the gatekeeper. This trend once again signalled the rise of blog triumphalism and the demise of media professionalism.

Despite the benefits, as a cautious optimist, Case Sunstein contended that the we should be vigilant to the "Daily Me" and coined the famous analogy— "information cocoon"— to warn that the "Daily Me" form of ICTs will inevitably induce groupthinks and polarization (Sunstein, 2006). Identified with the concept of the information cocoon, terms like the echo chamber effect and the filter bubbles are also frequently mentioned in the literature on related topics such as half-truths, fake news, anti-intellectualism, and populism on social media (Cinelli et al., 2021; Pariser, 2011b).

While the literature supporting the existence of information cocoons on social media is sufficient, there is a gap in connecting the underlying reasons, particularly human beings' inherent behavioural and cognitive characteristics, with the fact that social media is more prone to breed cocoons and thereby reshape brains. These will be discussed below.

**1.1 System 1 & 2: Addiction & Impatience**

Using Daniel Kahneman's classification, a dual process model is responsible for different patterns of thinking and decision-making (Kahneman, 2017). System 1 is an automatic mode, in charge of fast, instinctive, associative, and subconscious thinking, inclined to seek immediate rewards and make impulsive decisions. In contrast, System 2 is a reflective mode, in command of slow, effortful, rational, logical, and self-aware thinking, capable of calculating cost-benefits and thus making long-term decisions.

Although it appears that a brain operating on system 1 and indulging in immediate gratification mirrors the concepts discussed in *Amusing Ourselves to Death* (Postman, 2005), it is not necessarily safe to argue that system 2 is superior to system 1. The behavioural and cognitive studies in the context of biological evolution have proved that system 1 is far from a flaw, but rather a necessary feature for survival (Marg, 1995).

Specifically, our reptilian and mammalian brains, just as those of the animals, are responsible for instantaneous and spontaneous responses to the external environment in the case of danger-avoiding or reward-seeking for the purpose of maximizing the probability of survival. Instead of a defect, system 1 offers efficient, habitual shortcuts that allow us to maintain the basic life activities using only minimal efforts and energies (Kahneman, 2017). Nonetheless, system 2 deserves our respect since it is the cerebral cortex that distinguishes modern man' brain from those of animals and even that of Neanderthal man (Haldane, 2015).

In terms of media consumption behaviours, traditional media such as books, newspapers,

and TVs usually require audience's long-time attention and patience to capture the content (the signifier) and even certain level of literacy or numeracy to fully comprehend the meanings (the signified), in which case System 2 is called to work effortfully and deliberately. On the contrary, when we are surfing on the social media platforms and swiping the screen to ingest the memes, short videos and brief articles, System 1 runs quite effortlessly and this thinking pattern rewards us immediately via a dopamine mechanism (Macit et al., 2018), thus making social media more attractive and addictive than its traditional counterpart.

Because an individual's daily media consumption time is limited, there is a trade-off between deep reading (System 2) and fast reading (System 1). As a result, social media tend to win competition for users' attention due to its capacity to appeal to System 1 while filling time slots that would otherwise be reserved for System 2. And, as researchers have warned, this trade-off could reshape our minds by rewiring the brains and redistributing the relative weights of System 1 & 2, resulting in more addicted audience with impatient minds.

**1.2 Confirmation Bias, Hyper-sociability, and Division of Cognitive Labor**

Although System 1 explains why social media platforms are more appealing and addictive than conventional media, this pattern is insufficient to build an information cocoon. "Myside bias" is another component of the menu. Many psychological and cognitive research have found that facts alone always fail to persuade people to change their minds (Kolbert, 2017). One of the most fundamental reasons is that people tend to hold on to their initial beliefs and try to exaggerate the evidence in their favour while suppressing or disregarding dissenting viewpoints. This tendency is referred to as "myside bias" or "confirmation bias".

Similar to the observation that System 1 is not a design flaw of our brains, neither is myside bias. If logical reasoning and rational thinking based on system 2 are designed for sound judgement which benefits individual survival and social growth, the fact that both impartial reason and myside bias survived natural selection, as Mercier and Sperber argue, proves that this cognitive bias must have some adaptive function, which is related to "hyper-sociability" and "division of cognitive labour" (Mercier & Sperber, 2011).

Applying the "black box" theory in the contexts of science and technology, incomplete understanding of and directly using previous knowledges along with division of labour could empower progress to a large extent (Rosenberg & Nathan, 1982). By allowing successors to stand on the shoulder of predecessors and innovate further without mastering all previous principles and thus saving effort and resources, this phenomenon is also referred to as rational ignorance (Caplan, 2001).

Likewise, as the population grows and the society expands, it become harder for an individual to fully understand the complicated social network. In order to maintain communal trust and cooperation, it is reasonable to suppose that some extent of lop-sidedness and blindness is more useful than entirely rational reasoning (Kolbert, 2017). Moreover, under the circumstances that we are living in a hyper-connected and international society, we are confronted with more complex issues and wicked problems our forebears didn't deal with, such as fabricated studies and fake news. "This is one of many cases in which the environment changed too quickly for natural selection to catch up." (Mercier & Sperber, 2011)

In this way, myside bias provides both comfort zones for individuals to be satisfied with their situations and buffer zones for communities to preserve the peaceful and collaborative conditions. Apart from the data and information that could also be found on traditional media, these zones are exactly the unique goods and services that social media are catering to us. While the "division of cognitive labour" has benefited our hyper-society in many aspects, where it gets us into trouble is in the political domain (Fernbach et al., 2013), as shown by the rising populism and short-termism reflected in Brexit and Trump voters.

**1.3 Agenda-setting, Selective Exposure, and Framing**

The previous two sections illustrated the inherent characteristics of human cognition that contribute to the formation of information cocoons on social media. In terms of whether a form of media is more likely to produce cocoons, from the perspective of communication theories, this section will address the features of recommendation algorithms that distinguish social media from conventional media as well as obsolete digital platforms that do not use algorithms.

In the literature, researchers have summarized two levels of agenda-setting (Ghanem, 1997). The first level "focused on the relative salience (perceived importance) of issues or subjects" while the second level investigates the relative salience of qualities within issues (Weaver, 2007). Balmas and Sheafer (2010) added that the first level of agenda-setting focuses on the media's role in telling us "what to think about", whereas the second level telling us "how to think about". The concept of framing or attribute agenda-setting is frequently associated with second level (Ghanem, 1997; McCombs, 1997; Takeshita, 1997). When reporting a particular topic, framing assumes that the news media would emphasize some qualities (e.g., sub-issues or facets) while ignoring others, influencing how various audiences view the same issue, which is also known as the selective exposure (Stroud, 2008, 2010).

During the era of traditional media, professional journalists were educated and trained to report from various aspects based on impartial viewpoints (Schudson & Anderson, 2009). This orientation recalled the first level of agenda-setting and focused on informing the audience "what to think about" while avoiding imposing predispositions or guiding biases. However, in the age of social media, journalist professionalism is dead and grassroot "producers" make up the mainstream of media ecology (Weimann & Brosius, 2017).

Without editing, proofreading and gatekeeping, biased and emotional information along with fake news become the mainstream. Particularly, with the help of recommendation algorithms, similar biases and relevant emotions are often clustered, reducing the ideological diversity of a certain information stream while facilitating the formation of information cocoons—telling people "how to think about" while catering to the system 1. Therefore, identified with the second-level agenda-setting, the algorithm-based social media are more likely to produce information cocoons that could further rewire our brains.

## 2. Policy Problems: Information cocoons as mind-narrowing model

**Public Goods, Market Failures, and Externalities**

Despite the negative impacts of information cocoons on social media, some may argue that individuals should have their own rights to choose what to read and how to think. Why should we be worried about this personal-choice issue and even come up with policy interventions?

The most straightforward answer to this question could be the standard economic theories. In this way, information diversity (heterogeneity) and information cocoons (homogeneity) could be respectively regarded as public goods and market failures. And thus, as the economists put, negative externalities would be generated and ultimately undermine the social welfares. The following sections will introduce cases of impacted social welfares and explain details of the self-perpetuating failures to provide rationales for government interventions.

### 2.1 Bias & Noise in Judgement

Daniel Kahneman and Cass Sunstein have both dedicated a lot in studying how behavioural and cognitive "biases" would affect human judgement, which could be broken down into two processes—information-collecting and decision-making. In a co-authored book, *Noise: A Flaw in Human Judgment* (2021) they further argue that, apart from biases, "noises" constitute another behavioural factor that deserve our attention.

On the one hand, "biased" judgment causes systematic injustice such as the pervasive discrimination in the US justice systems where some judges tent to determine that defendants with certain traits are more likely to jump bail, which has been proved inaccurate (Kleinberg et al., 2018). On the other hand, "noisy" judgement causes occasional injustice due to random and unpredictable factors like timing, weather, and mood.

Kahneman, D., Sibony, O., & Sunstein, C. R. (2021) take fingerprint examination as an example to caution that even professionalism is neither noise-proof nor flawless in decision-making. We probably think of fingerprint comparison as a straightforward, mechanical and easily automated task. However, the latent prints collected from a crime scene are often partial, unclear, smudged, or distorted. In this regard, deciding whether they match a suspect's exemplar prints requires an expert's discretionary judgment. And the research findings suggest that even a tiny noise, such as timing, emotion, and anecdote, could significantly shift an expert's judgment, thus causing innocent people being mistakenly defined as a suspect.

To overcome biases and noises in decision-making, Kahneman, D., Sibony, O., & Sunstein, C. R. (2021) claim that even the naïve statistics and mathematics models (e.g., linear regression) would outperform human's judgment in a wide range of tasks, not to mention those state-of-the-art machine learning models. For instance, AI has proved to be more accurate in predicting what kinds of defendants are more likely to jump bail and which types of civil servants are more appropriate to be recruited or promoted. In this regard, it seems that algorithms could be more beneficial to social welfare than what have been argued in terms of information cocoons. However, a surprising and remarkable point worth noting is that, noises, in contrast to biases, can play diametrically opposed roles in the dual processes of judgement—information-collecting and decision-making. Simply put, noises are detrimental for decision-making but advantageous to information-collecting.

As argued in *Homo Deus* (2015), organisms are analogous to algorithms (Harari, 2016). Compared with computers, the most salient difference is that organisms are more subjective to noises (e.g., emotions and preferences). But if systematic methods of evaluation are applied, human judgement would also become more computational. For example, to determine which candidate is more suitable for a job position, the judges would first assign certain weights to desired variables, such as 50% for education experiences, 30% for professional skills, and 20% for personality traits. And then various aspects of data would be extracted from the candidate's resume and interview and would be scored in accordance with the predefined variables. Finally, weights and scores would be used together to calculate the final grade of the candidate. In this way, the evaluation and selection are structured with robustness, with the ranking of the final grades providing a reliably noise-less list for deciding the final candidate.

In the job-seeking example, the weight-assigning, performance-scoring, and grade-calculating are all components of decision-making, in which noises (e.g., arbitrary weighting or randomly discretional scoring) should be avoided. In contrast, for information-collecting, it is favourable to extract data and collect evidence from sources as diverse (noisy) as possible to take into accounts multiple facets of the certain issue and to avoid myside bias. Otherwise, even AI would make biased and even racist decisions based on incomplete or manipulated dataset (O'neil, 2016; Zou & Schiebinger, 2018). In a nutshell, the dichotomy between the roles of noise also unveils a crucial fact that, given bias-free and noise-free decision-making process, even algorithms cannot perform better than human judges and thus eventually output biased outcomes if the input data are noise-less. In other words, noisy information-collecting help reduce biases in judgement.

In the context of our daily life, to feel confident in a judgment, two conditions must be satisfied: the story must be comprehensively coherent, and there must be no attractive alternative interpretations (Kahneman et al., 2021). In the scenario of information cocoon, it appears that only like-minded people clustered and homogeneous opinions echo in a noise-less chamber. As such, comprehensive coherence is achieved by filtering out or supressing alternative evidence—a well-documented process in perception that could induce the so-called illusion of agreement.

On the contrary, the true experts on judgment have the "virtue of humility" and "actively open-minded thinking" during information-collecting. They know not only why their explanatory story is correct but also why other stories are wrong (Kahneman et al., 2021). They are aware that their judgment is a work in progress and yearn to be corrected. So, they actively search for information that contradicts the pre-existing hypotheses. They update their forecasts, like a Bayesian machine, in response to new information and thus adjust the weights of different pieces of evidence, despite the risk of being shameful or criticized.

To characterize the thinking style of these experts, Tetlock (2017) borrows the phrase "perpetual beta" from computer programming, which describes a software that is not meant to be released in a final version but is endlessly used, analysed, and improved. As he puts it, "What makes them so good is the hard work of research, the careful thought and self-criticism, the gathering and synthesizing of other perspectives, the granular judgments and relentless updating." They prefer a particular cycle of thinking: "try, fail, analyse, adjust, try again."

However, being such a "perpetual beta" requires effortful operation of System 2, which rarely happens naturally. Even though "passively open-minded thinking" is also beneficial to some extent, the noise-free algorithms reduce the information heterogeneity on an individual's information stream on social media, thus raising the possibility that the users would be confined by the information cocoons and would become habitual System 1-users. One study states that the scores indicating whether people will fall for "fake news" are associated with how much people use their smartphones (Pennycook et al., 2015), reflecting the propensity of social media audience to use impulsive (System 1) versus reflective (System 2) thinking patterns.

At a micro-level, the information cocoons as the mind-narrowing model would gradually intensify the System 1 style of thinking, inculcate biases, and deceive people into fake news, thus leading to wrong judgement within contexts of family and workplaces. At a macro-level, this model would incur cyberbullying, cause social injustice, and even undermine the root of democracy while inciting populism at a large scale via voting mechanisms.

**2.2 Populism: Cultural backlash in Brexit & Trump**

As two famous black swan events, the 2016 Brexit referendum in the United Kingdom and the 2016 presidential election in the United States both reflect the phenomenon of populism. Among various determinants in populism, some studies argue that "cultural backlash" (Inglehart & Norris, 2016) is more significant than social and economic variables.

Furthermore, considering that social media has become the most popular and powerful weapon in the culture industry as to 2016, taking the roles of producing and distributing cultural meanings, as well as reproducing and reinforcing dominant ideologies (Horkheimer & Adorno, 2017), "cultural backlash" as a factor is also highly related to information cocoons caused by social media algorithms. This phenomenon is also referred to as "filter bubbles" in related studies (DiFranzo & Gloria-Garcia, 2017; Geschke et al., 2019; Napoli, 2018), in which empirical findings suggest that social media algorithms on Twitter and Facebook help create ideological "bubbles" by clustering congenial political ideas while filtering discordant voices on individual's information stream, thus intensifying the confirmation biases of the constituents, contributing to political polarization, and resulting in the observed voting outcomes.

Ever since the establishment of the European Economic Community (EEC) in 1957, the issues regarding Europe Integration and UK's membership have triggered controversial discussions and thus formed a Eurosceptic climate nationwide, which eventually contributing to the emergence of the Brexit agenda (Childs, 2012). The Brexit Leave campaign and UKIP (UK Independence Party) rhetoric harkened back nostalgically to a time before joining the EU, the so-called "glory days of empire", when the society was predominately white Anglo-Saxon, manufacturing factories and extracting industries (producing steel, coal, cars) still provided well-paying and secure jobs for unionized manual workers.

Likewise, Donald Trump's slogan "Make America Great Again (MAGA)" and his rejection of "political correctness" appealed nostalgically to a mythical "golden past", especially for the elderly white men who embrace the conventional sex roles and patrimonial power relationships. Similar messages can be heard in the rhetoric of Marine Le Pen, Geert Wilders, and other populist leaders (Inglehart & Norris, 2016). This nostalgia is most likely to appeal to older citizens who have seen social changes erode their "cultural predominance" and threaten their core social values, provoking a response expressing anger, resentment, and political disaffection as the "cultural backlash".

Given the consequences, however, the Brexit referendum and Trump Administration proved to be ill-advised decisions, giving rise to racism, nationalism, slower globalisation, premature de-industrialisation, aggravated industrial hollowing and inequality (Inglehart & Norris, 2016). The complexity of the economic and societal implications was far beyond normal citizens' knowledge and political awareness. Additionally, given the manipulated public opinions and media agendas, the political campaigns equipped with framing strategies in Brexit and Trump successfully played important roles in reinforcing pre-existing attitudes and swaying the undecided groups, leading to the final winning by a narrow electoral margin.

First, ***"Reasoning Devices"*** provide evidence to logically support the framing (Entman, 2002). Taking the Brexit, the leave campaign used the case of "Abu Hamza's daughter-in-law" to argue that the overruling EU court and EU-led policies had undermined UK's sovereignty and were posing threats to UK's national security. Plus, they also took Turkey as an example to illustrate the potential risks of European expansion resulting in problems on national job markets and public services (Lundblad, 2017). Although the cases that Leave Campaign used were factual information, they somewhat distorted the information in favour of their position and propagated half-truths by concealing the parts that would have weakened their arguments.

Specifically, when talking about the topic of EU membership price, the Leave Campaigner declared that "the EU cost the UK £350 million per week", one of the most popular catchy phrases (memorable Vocabulary) during the campaign, ignoring the fact of rebates and benefits to the UK. In effect, however, the actual net contribution was about £140 million per week. In addition, "take back control" was another famous slogan aiming at the sovereignty issue by arising citizen's memory of the UK's national identity in terms of imperialistic heritage and its legacy as a world leader in British history.

By establishing a causal linkage between EU membership and social problems in the UK, though not logically robust, these communication tricks successfully instilled misperception into citizens and to a large extent helped the Leave Campaign complete the ***"Moral***

*Evaluation"*—it is the EU that should be blamed as the root cause. Then, by deceptively stating that there were "no alternatives", the Leave Campaign finished the ***"Remedy Suggestion"***, in which "Vote Leave" is the only option (Lundblad, 2017).

It turned out that Leave Campaign's framing played a significant role in determining the outcome (Gavin, 2018; Weimann & Brosius, 2017). One convincing evidence is that 1.2 million people who voted Leave regretted their decision after the election (Lalić-Krstin & Silaški, 2018), indicating that the opinions of the undecided and/or unstable groups were successfully shifted by the framing strategies. By contrast, the Remain camp's publicity was unattractive and "uninspiring in the extreme" (Cassidy, 2016). According to statistics on Circulation and Reach, the Pro Leave tabloids (e.g., The Sun, Daily Mail) reached most readers with the highest circulation figures (48%), whereas Pro remain articles written by professional broadsheet (e.g., Financial Times, The Guardian) accounted for just 22% (Levy et al., 2016). In other words, the pro-leave campaigns leveraged System 1 among readers to attract attentions and win supports while the pro-remain camps lost the competition for attention only with the hope to arouse rational thinking and long-term decision-making provided by System 2. It is reasonable to presume that, during the campaigns, social media as the major source of news with recommendation algorithms, further exacerbated the imbalance of information distribution and attention allocation.

In retrospect, the remain camp could have leveraged big data technology and social media algorithms to affect the audience' information-collecting and enhance their framing strength, thus not only reinforcing the pro-remain attitudes but also shifting the opinions of the unstable but crucial constituents, which might have led to a global political dynamic that is more favourable to international cooperation, R&D innovation, and sustainable development.

**2.3 Short-termism, Innovation and Sustainability: Think slow, grow fast**

<u>2.3.1 Humanistic capitals for long-term social growth</u>

In the whole history of modern man, besides the triumph over the Neanderthal man, the most remarkable leap in social growth happened after the first industrial revolution, which gave rise to the explosion of sciences and technologies as well as the sharp increase in GDPs and social welfares globally (Haldane, 2015). However, apart from the physical capital from a science and technology perspective, human, intellectual, social, and infrastructure (institutional) capitals reflected in the fields of humanities and social sciences are also pre-requisite conditions for a society to innovate and grow sustainably (Sachs, 2014).

By examining the historical data and underlying tendencies related to human development and social growth, researchers contend that the increasing use of system 2 (i.e., the slow-thinking, reflective, patient part of the brain) has contributed significantly to the accumulation of those four humanistic capitals before the industrial revolutions (Haldane, 2015). From a financial investment perspective, the investor who prefers long-term rewards would restrain her short-term impulsion to consume the money immediately but rather invest in pursuit of more rewarding outcomes in the far future.

However, just as stock investment, this so-called long-term reward is never guaranteed in other investing activities such as education, healthy diet, and R&D. So, the risks and uncertainties in long-term investment are the barriers to System 2. Especially when the profitable investment opportunities are become rarer, it appears that since kick-off the information revolution, System 1 (i.e., the fast-thinking, reflexive, impatient part of the brain) has re-started to expand its influence (Haldane, 2015). If so, that would tend to raise societal levels of impatient time-preference, foster shorter-term decision-making, and slow the accumulation of all types of capital for long-term growth. In a word, fast thinking could make for slow(er) growth. And a society teeming with information cocoons as the mind-narrowing model on social media exactly echoes this dreadful prospect in a self-perpetuating way.

2.3.2 Complex Systems: micro-diversity, heterogeneity, and positive deviance in network

The assumption underpinning neo-institutional economics is rational instrumentalism (Berry et al., 2004). And under the circumstances that this assumption fits, the wisdom and intelligence of the individuals who have the common goals in organizations could be gathered through the network and then used rationally to overcome challenges and reduce transaction costs that otherwise would be prohibitive. This phenomenon is referred to as collective intelligence (Sunstein, 2006). And the most cited examples are the establishments of open-source communities—Wikipedia and GitHub.

In fact, however, there are so rare and precious scenarios in which we could naively rely on the rationality assumption and expect the benefits of the network. For the most part of the reality, we should also think more about the dark side of networks as well as the assumption behind the social network analysis—people's behaviours and performances are subject to context and socialization. "Groupthink" is such a detrimental phenomenon that is common in public or private organizations where like-minded voices are encouraged while dissent is suppressed, whether because of the organizational culture or the existence of a domineering leader (Janis, 1983; Sunstein, 2006). In this regard, Granovetter not only emphasized the contextual embeddedness assumption but also contended that overly "strong ties" in networks are apt to nurture "groupthink" (Granovetter, 1973), a direct opposite of collective intelligence. In a network where most members are committed to preserving existing norms and loyalties while rejecting competing opinions, there would be high homogeneity of beliefs and attitudes, framing this type of network an information cocoon or echo chamber.

Therefore, by focusing on the contextual, psychological, and behavioural aspects of networks, it is meaningful to figure out which features and configurations are significant and how to design the network structures in order to mitigate the negative effects while maximizing advantages. Complexity science also argues for micro-diversity at individual level as well as information heterogeneity at network level for the purpose of identifying and amplifying innovative ideas (Goldstein et al., 2010). It is hard, if not impossible for same old opinions to inspire novelties. So, complexity science studies encourage departure from the original path of thinking and learning, which is referred to as "positive deviance".

And because of the features of network, a tiny improvement in only a small number of nodes could potentially spark considerable changes in the whole dynamic of network. As opposed to a linear or hierarchical structure, where the information can only proceed in one direction strictly layer by layer, the nonlinear structure in a network where people can be linked freely means that every two pairs of nodes in the network can directly and interactively communicate and share information with each other. And thus, through simple mathematics, we could derive an equation for the number of links given a network with n nodes—Number of Links = $n(n-1)$. Obviously, with a linear increase of n, the number of links would rise non-proportionally (close to exponentially) higher. And this phenomenon is exactly the nonlinear effect that could inspire and encourage policy-makers to improve the information diversity and interaction resonance, thereby potentially enhancing innovations and social growth.

## 3. Policy Analysis: Information sidewalks as diversity-boosting mechanism

### 3.1 Policy Objective, Stakeholders, Criteria, and Alternatives,

3.1.1 Policy Objective

Based on the policy contexts and policy problems discussed above, the policy objective in terms of this issue is to regulate social media algorithms in order to improve the diversity of individual's information consumption, with a long-term goal of encouraging the use of System 2 in judgement at micro-level and enhancing sustainable social growth at macro-level.

To achieve this objective, the inventor of the term "information cocoons", Cass Sunstein had also employed a "sidewalk" analogy to depict the diversity-boosting mechanism (Geng, 2006; Sunstein, 2006). As opposed to an echo chamber in which expected ideas embrace each other with rare surprises, the sidewalks are where we will accidently come across people from different backgrounds with a wide variety of ideologies. On the intersection of diversified discourses, innovations would happen just as serendipitous discoveries emerged from interdisciplinary studies (Darbellay et al., 2014).

3.1.2 Stakeholders and Criteria

Before proposing policy alternatives to replace the information cocoon model with the information sidewalk mechanism, we need to identify the stakeholders and criteria involved in this issue, which will be then used to analyse stakeholder positions and project policy outcomes as shown in the appendix (Table 1 and Table 2).

From the demand side of the market, the audience is the stakeholder who directly consume information provided by social media platforms. From the supply side of the market, the social media companies, their financial investors, and potential advertisers constitute the group of stakeholders whose monetary interests would be directly affected by policies. Specifically, their profits almost exclusively derive from targeted advertising (Fuchs, 2018). And the profitability further influences the market evaluation of companies that is the core concern of the investors.

From the perspective outside the market, the government and academia are the stakeholders whose interests could be indirectly affected by this issue. And they both have external powers to impact the dynamics and change the status quo of the social media market. Besides, during the policy formulation and implementation, academia can not only provide theories and methodologies for the government, but also enhance the reputation and social trust of the paternalist actions and regulatory operations among the citizens (Wu et al., 2017).

Based on big data analysis of users' browsing history and demographic and geographic characteristics, social media companies classify target customers with tags. Then recommendation algorithms match and distribute certain advertisements to users whose statistics theoretically maximize the probability that the ads would be clicked (Knoll, 2016).

In terms of winning the bargaining power to attract more advertisers with higher quoted prices, the social media companies could maximize ads profits in two ways. First, aligned with attention economics, given limited time and scarcity of attention, companies attempt to attract users to stay on their platforms to kill time for as long as possible. And by building information cocoons, platforms provide comfort zones for users to amuse and enjoy themselves during the information consumption, thus exploiting users' time and attention. Second, based on the logic of recommendation algorithms, the more polarized and clustered the users are, the easier the algorithms could tag the potential consumers. In this sense, information cocoons also help purify target markets by simplifying users' data characteristics, which could thereafter help increase the clickthrough rate (CTR) with high validity and accuracy (Yang et al., 2016).

So, when considering policy alternatives, the resistance from the supply-side stakeholders is an important factor in designing the policy tools. Besides, due to their high resource mobilization capacities, the positions of these three stakeholders play significant roles in the stakeholder analysis (Table 1). Thereafter, four criteria—cost, effectiveness, political acceptability, and administrative operability—will be used to project the policy outcomes in the matrix (Table 2) to evaluate the alternatives.

### 3.1.3 Policy Alternatives

There are generally four categories of policy interventions—mandates (e.g., bans & mandatory standards), communication (information & education), standard economics (taxes & subsidies), and behavioural economics (bias & nudge). Considering the potentially strong backlash from the supply-side stakeholders in response to mandate coercion, this paper proposes three policy alternatives—Training, Taxing, Nudging—respectively derived from the latter three categories.

#### Alternative A: Training

Speaking of how to improve judgement in terms of managing bias and noise, Kahneman, D., Sibony, O., & Sunstein, C. R. (2021) suggest that training can make a difference. There is some evidence that the abovementioned "actively open-minded thinking" is a teachable skill. This finding also echoes that, given a baseline of the using pattern of system 1 & 2, an individual's preference of using either one can be significantly affected by external factors (Daniel, 2017; Sunstein, 2014). These facts justify training as the policy alternative A.

For example, apart from literacy and numeracy, information-literacy could be raised as another important capacity for modern citizens to adapt themselves to the data flood in the fast-changing information era. Specifically, in the curriculums of various levels of school, the governments could design and add courses that teach students to acknowledge behaviour and cognitive biases, to identify half-truths and fake news on social media, and to become the "perpetual-beta" type of information consumer in judgement tasks. Although this policy alternative is not difficult to operate due to its clear scope and division of labour, it would take long for the effects to become salient and the originally imbalanced distribution of education resources would also cause uneven effectiveness of this policy.

In terms of political acceptability, because this alternative only takes effects on the part of the citizens, and the resulting effects are not easy to be detected, the supply-side stakeholders will not show significant positions regarding this issue, as shown by the zero scores in Table 1. Likewise, since training could be a mild and long-term process integrated into the conventional education experiences, it is not easy for the audience to perceive this policy alternative and thus they will also express few substantial attitudes toward this alternative.

However, as the major payer of public education, the governmental departments would constitute the greatest opponent of this policy alternative. On the one hand, the added courses and teachers would increase the total costs that the governments spend on the public education sectors. On the other hand, given the relatively fixed budgets, the departments other than the education sectors would resist this plan in order to strive for their own budget allocation. On the contrary, as a co-related member of the education sectors, the academia as the stakeholder would become the major proponent of this alternative, also because training reflects academia's intellectualism philosophy that citizens with independent and critical thinking would be beneficial to the long-term and sustainable social growth.

<u>Alternative B: Taxing</u>

As the main source of government revenue, taxation is an effective policy tool that generates economic incentives for stakeholders to change behaviours while transferring incomes from the winners to the groups in need. Considering that social media companies are the monopolistic winners who make profits based on the exploitation of users' data while producing negative externalities, there are rationales for the governments to regulate the market.

In recent years, there has been a worldwide tendency for data regulations toward IT companies (Economics, 2020). As the "Data Security Law" and the "Personal Information Protection Law" successively took effect in the second half of 2021, China's governments are tightening up controls on data-driven companies such as Tencent, Alibaba, and DiDi. Likewise, in 2020, their overseas counterparts, Facebook, Amazon, and Uber had already been regulated by privacy laws in force, such as EU's General Data Protection Regulation (GDPR), as well as the US's California Consumer Privacy Act (CCPA).

Based on the categories of goods in standard economics, there are three metaphors for data economy—oil (private goods), sunlight (public goods), and infrastructure (club goods) (Economics, 2020). In the early days, data are predominantly treated as oil, indicating that whoever extracts them owns them. This mode of data economy benefits the digital platforms to the largest extent, generating multiple Internet giants globally. However, the difficulties in defining the property rights of user-generated data made it unfair for the users to protect and benefit from their own data rights. Meanwhile, digital giants tend to become monopolies, exploiting data values for marketing and R&D while occasionally breaching data privacy and information security, as shown by the Cambridge Analytica scandal (Hinds et al., 2020).

Therefore, the oil model is flawed and problematic for social welfare. Especially when realizing that a remarkable nature of data is duplicability, it is obvious that this "non-rivalrous" characteristic echoes the public and club goods. So, these two models are plausibly more applicable in the context of data economy. For example, in some European countries, open-data movements and projects are popular. In this scenario, business and healthcare data are distributed publicly for free like the sunshine, for the purpose of maximizing the data utility and social welfare.

Nevertheless, this wonderful data utopia would be extremely hard to be realized. The most significant dilemma is that if data are completely "non-excludable", the quality of data is hard to be guaranteed. Data per se are somewhat like oil because they need to be refined to be useful. In the data industry, only after being "organized", "cleansed" and "tagged" will statistical patterns be explored from data sets (Economics, 2020). In this regard, if data are treated as public goods, although everybody can get access to data without any cost just like enjoying sunbathing, it is difficult for data sets to be exploited effectively. And the utopia would end up being teeming with low-quality data sets like pollutions as another negative externality.

Therefore, it seems that the infrastructure (club goods) model would be more appropriate. In this model, each club member must pay for the club fee (in the form of taxation) to get the right to exploit data. In this way, data are "non-rivalrous" but "excludable", which addresses the trade-off between data privacy and data value. So, the policy alternative B—taxing—is the key method to ensure that the data economy runs in this way.

However, despite that taxing are welcome in the positions of governments, academia, and audience, this alternative will incite the greatest resistance among the supply-side stakeholders, resulting in the lowest political acceptability and administrative operability. Especially in the democratic countries and regions where hard paternalism is rarely successfully operated, this policy alternative is hardly feasible.

<u>Alternative C: Nudging</u>

After examining the trade-offs between policy criteria demonstrated by policy alternative A and B, nudge, as a novel policy tool based on behavioural economics, could provide soft paternalism that balance performance in each criterion while satisfying every stakeholder. The details justifying nudge as the policy alternative C will be discussed in the next section.

### 3.2 Nudges: The Red pill & Blue pill metaphor

In *The Matrix* (Wachowski et al., 1999), the terms "red pill" and "blue pill" refer to a choice between the willingness to learn a potentially unsettling or life-changing truth by taking the red pill or remaining in contented ignorance with the blue pill. And this metaphor seems highly identified with the topic of information cocoon. Specifically, if a social media user is willing to choose to indulge in the cocoons, the corresponding algorithm could be defined as the "blue pill" mode. In contrast, in the "red pill" mode, the algorithms deliberately add diverse topics and dissenting opinions like building information sidewalks in individual's data stream.

Instead of mandatorily requiring the social media to be in either mode pre-dominantly, the policy alternative C—nudging—proposes to provide users with free choice, just as Neo's chance to choose in the film, along with a default choice being the "blue pill" mode. Rationales for and benefits of this alternative will be discussed in detail in the following sections.

3.2.1 Soft Paternalism: choice architecture is inevitable

In terms of the degree of government intervention, "Communication" and "Nudge" can be classified as soft paternalism while "Coercion" and "Incentives" as hard paternalism. But in the view of liberalists and free-market advocates, both types of paternalism are not justified.

In Mill's "Harm Principle", it is only when "there is definite damage, or a definite risk of damage, either to an individual or to the public," that "the case is taken out of the province of liberty and placed in that of morality or law." (Mill, 1975). In Mill's view, the problem with outsiders, including government officials, is that they lack the necessary information. Likewise, Friedrich Hayek emphasized the dispersed nature of human knowledge and the informational advantages of markets, incorporating that dispersed knowledge, over even the most intelligent and well-motivated planners (Hayek, 2009).

The counterarguments to the anti-paternalists are as follows. Despite that standard economic theory argues that people will reasonably analyse conflicting evidence and rationally strike balances between the present and the future, in practice, people tend to fall into behavioural errors such as confirmation bias as discussed before as well as present bias, in which people procrastinate or neglect to take steps that impose small, short-term costs but would produce large, long-term gains. While System 2 considers the long term, System 1 is myopic. Furthermore, free markets may also reward sellers who attempt to exploit human errors and market failures. In the light of Gresham's law, those who do not exploit human errors will be seriously punished by market forces, simply because their competitors are profiting from doing so—racing to the bottom (Sunstein, 2014), as demonstrated by the pervasive principal-agency problems in the 2008 subprime mortgage crisis.

Instead of regarding this issue as an either-or situation, it makes far more sense to respect both arguments and strike a balance between the extremes. Just as considering that although System 2 provides advanced functions, we also acknowledge that System 1 is not a flaw. So, it is more reasonable to recognize that people display bounded rationality rather than accusing them of "irrationality". For many purposes, bounded rationality is even beneficial in terms of making us emotional humans rather than robots as well as producing outcomes that are equal to or perhaps even better than effortfully analyzing costs and benefits (Sunstein, 2014). In this sense, nudge as a policy tool could also effectively alter human behaviours by leveraging human biases and directing them into the desirable directions.

More importantly, the common knowledge we should extract from both arguments is that, whether it is a free market or a regulated market, the framing of choices (i.e., choice architecture) is inevitable (Sunstein, 2014). The only difference exists between the choice architects—private or public sectors. Therefore, respecting the benefits of personal liberty and free markets, meanwhile noticing the alarms rung by behavioural errors and market failures, it is justifiable to suggest that soft paternalism, especially nudges, could be the most appropriate policy alternative that strikes the balance between effectiveness and acceptability (Table 2).

3.2.2 Choice-preserving: respecting the freedom of choice

Some welfarist contend that freedom of choice, per se, is welfare (Sunstein, 2014). In addition, people are highly diverse in terms of their tastes, their values, and their situations. So, a one-form-fit-all policy is not feasible in this setting and it is endlessly controversial to argue whether collective social welfare outweigh the total of those of individuals.

To avoid this dilemma while still attempting to mitigate the collective action problem related to information cocoons, policy alternative C—nudging—is the mildest and most choice-preserving forms of intervention, understood as initiatives that maintain freedom of choice while also steering people's decisions in the right direction (as judged by people themselves).

### 3.2.3 Power of Default: leveraging the inertia bias

It is also worth noticing that in some contexts, people do not enjoy freedom of choice and would much prefer not to have to spend time on the question at all (Sunstein, 2014). Especially in complex and unfamiliar domains, active choosing can be a burden rather than a benefit. So, in terms of the "blue pill" or "red pill" question, the users who are not familiar with this concept and its background story would feel frustrated to make a decision.

In this regard, inertia is often exceedingly powerful (Kahneman, 2017; Sunstein, 2014) and it helps account for the power of default rules when it comes to nudge, which establish what happens if people do nothing. Often nothing is exactly what people will do, so the default rule tends to stick, whether it involves saving for retirement, personal privacy, or clean energy. Therefore, it is reasonable for the policy-makers to leverage the inertia bias and set "blue pill" as the default rule in favour of their desired direction, though this setting would undoubtedly provoke the resistance from the supply-side stakeholders to some extent (Table 1).

### 3.2.4 Social Norms: leveraging herding and groupthink

"Thinking socially" also contributes to the success of nudging. Since humans are social creatures surrounded by social meanings, the audience is broadly subject to herding and groupthink to secure the membership of certain communities or to avoid being falling behind emerging rituals in response to an opinion climate or a new cult created by the mass media (Blumer, 1986; Lippmann, 1929). This illustrates that social norms and peer influence are also determinants of shaping people's consciousness and shifting behavioural habits. For example, by tagging a tiny "pill" mark on user's ID or profile picture, the negative connotation conveyed by the "blue pill" will help promote the trend of choosing the "red pill" mode on the Internet.

### 3.3 Caveat: Government Failures

3.3.1 Disguised Censorship & Regulatory Arbitrage

It is undeniable that government officials can also err, and their errors may be especially, even uniquely, damaging. The reasons for government failures are two-fold. First, as stated by the public choice theory: The judgments of officials about welfare may be influenced by the interests of powerful private groups (Sunstein, 2014). At some times and places, official judgments have been distorted because of the power of such groups. Second, even if they are well motivated, officials are human too, and there is no reason to think that they are immune from the kinds of biases that affect ordinary people (Sunstein, 2014). So, even the most benign paternalists can go badly wrong, and some paternalists are far from benign.

The most salient government failure that would arise is that the information free market would be manipulated accidentally or deliberately by the policy interventions in social media. For example, if the government departments make the use of this window opportunity to participate in the design and management of algorithms, there would potentially be risks that the governments implant specially-designed programs serving as a way of disguised censorship. Even if there is no intentional manipulation, there would also be inadvertent mistakes such as the regulatory arbitrage, in which case only parts of the stakeholders (social media companies) are covered by the policy, leaving the rooms for the rest to circumvent the regulations and continue to exploit the market. In this way, the policy undermines the equality of the market by inducing extra negative externalities.

### 3.3.2 Ethical Discourse: the prospect of sound policy-making

As suggested by the literature on digital governance, it is important for the government to realize that a well-established ethical discourse (Haque, 2003) between the private sectors (e.g., social media companies), the public sectors, the academia, and the citizens, is crucial for the policies on digital governance to be both ethically justiciable and technical reasonable. In this case, no matter which policy alternative is used, open communication, transparent disclosure, and effective public participation are indispensable in the full life-cycle of policy-making.

## Recommendations and Limitations

With the goal of establishing "the information sidewalks as the public forum", this paper proposed three policy alternatives—Training, Taxing, Nudging—to fight information cocoons and foster a sustainable information environment that might benefits long-term social welfare.

By analyzing the stakeholder positions (Table 1) and criteria performances (Table 2), it reveals that policy alternative C—Nudging—would be the most effective one primarily because it directly reforms the recommendation algorithms. And although nudging would provoke resistance from supply-side stakeholders to some extent and thus become a litter inferior to alternative A—Training—in terms of political acceptability, it is also far more acceptable than alternative B—Taxing. Besides, alternative C far outperforms alternative A with respect to the burden on public budget. Finally, despite the highest profitability of alternative B, the lowest political acceptability and administrative operability would make it the least desirable and most impractical policy alternative especially in countries or regions where hard paternalism is normally unwelcome.

Even though taxation is easy to implement in areas where the government is powerful, it seems that taxing alone cannot directly address the information cocoon issue, the revenues derived from taxation could be transferred to training projects mentioned in policy alternative A or the other corrective projects such as subsidizing the media professionalism. Each policy tool has its own advantage to tackle a specific type of problem. So, it is reasonable for the policymakers to embrace a diverse set of policy tools, just as using a Swiss army knife, to leverage advantages of alternatives to deal with certain aspects of a complex issue (Low, 2020).

In conclusion, on one hand, this paper recommends that policy alternative C could be the most appropriate one to be generalized and employed in a wide range of jurisdictions, including those in which the hard paternalism normally faces the greatest political and social resistance. On the other hand, in areas where hard paternalism is also feasible, a combination of advantages of different policy alternatives could be used to target specific respects of the policy problem while offsetting each other's shortcomings. Specifically, the governments could use Taxing (alternative A) as the financial source to support Training (alternative B) and as the bargaining chip in negotiations with supply-side stakeholders to promote Nudging (alternative C).

Due to the limits on time and effort, this paper did not delve into details of the theoretical principles and operational mechanisms of recommendation algorithms on social media. Neither did the paper try to investigate the information cocoon (filter bubble) phenomena through the lens of empirical or quantitative studies. Additionally, the performances of policy alternatives on each criterion are determined somewhat arbitrarily instead of basing on data analysis, which could undermine the accuracy and robustness of the policy analysis to some extent. Nonetheless, these limitations would also illuminate the research gaps and potential directions that further research could attempt to address.

# Appendix

## Table 1: Stakeholder Analysis

| Groups | Interest in issue | Resources available | Resource mobilization capacity | Position on Alternatives | | |
|---|---|---|---|---|---|---|
| | | | | A | B | C |
| Social Media Companies | Global Market share | Technology, market dominance | Very high | 0 | -3 | -1 |
| Investors | Return on investment (ROI) | Financial capitals | Very high | 0 | -3 | -1 |
| Advertisers | Clickthrough rate (CTR) | Advertising budget | Very high | 0 | -2 | -1 |
| Audience | User experience, data privacy, information relevance, comfort, joy, new knowledge | Consumer choice, complain, protest, campaign | Low to Medium | 0 | +1 | 0 |
| Governments | Data security and ownership, democracy, social growth, government revenue | Bureaucracy, political influence, policy tools, power of execution | Very high | -3 | +3 | +2 |
| Academia | Democracy, open-minded audience, long-term decision-making, sound judgement | Theories, academic influence, social reputation and trust | Medium to High | +2 | +2 | +2 |
| Total Scores | | | | -1 | -2 | +1 |

## Table 2: Alternative-Criterion Matrix

| Alternatives / Criteria | Alternative A Training | Alternative B Taxing | Alternative C Nudging |
|---|---|---|---|
| Cost | High | **Low** | Low to Medium |
| Effectiveness | Medium | Low to Medium | **Medium to High** |
| Political Acceptability | **High** | Low | Medium |
| Administrative Operability | Low to Medium | Low | **Medium to High** |

Note: Highlighted cells indicate the best performance on respective criteria.